\documentclass[12pt]{iopart}
\usepackage{psfig}

\begin{document}
\title[Nuclear-spin qubits interaction in mesoscopic wires and rings]{Nuclear-spin qubits interaction in mesoscopic wires and rings}
\author{{\it \ Yu.V. Pershin}$^{1,2,3}${\it , I.D.Vagner}$^{1,3,4}${\it \
and P.Wyder}$^{1}$}
\address{$^{1}${\small Grenoble High Magnetic Field Laboratory,}\\
Max-Planck-Institut f\"{u}r Festk\"{o}rperforschung and CNRS,\ \
\\ BP 166, F-38042 Grenoble Cedex 9, France,\\ {\small \
}$^{2}${\small B.I.Verkin Institute for Low Temperature Physics}\\
and Engineering,\ \\ 47 Lenin Avenue 310164 Kharkov,
Ukraine,{\small \ }\\ $^{3}${\small {Department of Physics,
Clarkson University, }}\\ Potsdam, New York 13699--5820, USA,\\
{\small \ }$^{4}${\small Research Center for Quantum Communication
Engineering, \\ Holon Academic Institute of Technology, \ }\\ 52
Golomb St., Holon 58102 Israel.}

\begin{abstract}
Theoretical study of the indirect coupling of nuclear spins
(qubits) embedded into a mesoscopic ring and in a finite length
quantum wire in a magnetic field is presented. It is found that
the hyperfine interaction, via the conduction electrons, between
nuclear spins exhibits sharp maxima as function of the magnetic
field and nuclear spin positions. This phenomenon can be used for
manipulation of qubits with almost atomic precision. Experimental
feasibility and implications for quantum logics devices is
discussed.

\end{abstract}

\section*{Introduction}

The possibility that a computer could be built employing the laws
of quantum mechanics has stimulated huge interest in searching for
useful algoritms and realizable physical implementations
\cite{Steane98}. There are currently many promising approaches to
quantum computation \cite{Fortschr}, the most promising solid
state approaches are based on the superconductivity \cite{Vion},
conduction electron spins \cite{dot} and nuclear spins \cite%
{PVK98,Kane98,Kane00,MPG01,SSV01,Privman02} as qubits. The
attraction of the approach of using nuclear spin in quantum
computation lies in the idea of incorporating nuclear spins into a
semiconductor device \cite{PVK98,Kane98,Kane00}. Using of the
nuclear spins incorporated in a heterostructure in the highly
nondissipative quantum Hall effect regime, as qubits \cite{PVK98},
are promising because they are extremely well isolated from their
environment and have a long decoherence time
\cite{MPV01,BMVW01,VM88}. The energy gap in the spectrum of two -
dimensional electrons in a strong magnetic field imposes severe
restriction on the flip-flop processes, since the electron Zeeman
energy is orders of magnitude larger than the nuclear one
\cite{VM88}. It follows that the nuclear spin imbedded in 2DES
under the QHE conditions is practically decoupled from the
conduction electron spins. As a result the nuclear spin relaxation
time has an activation behavior, i.e. is exponential in the
electron energy gap and inverse temperature \cite{VM88} and can be
manipulated by external magnetic field and sample parameters over
several orders of magnitude time interval.

A set of five essential criteria for the physical realization of a
quantum computer was formulated by DiVincenzo and coworkers
\cite{Divincz}. They are:

1. A scalable physical system with well characterized qubits. The
system should consist of a collection of independent subsystems
each with a two-dimensional Hilbert space, so-called quantum bits,
or qubits. Its physical parameters should be accurately known,
including the internal Hamiltonian of the qubit, the presence of
and couplings to other states of the qubit, the interactions with
other qubits, and the couplings to external fields that might be
used to manipulate the state of the qubit.

2. The ability to initialize the state of the qubits to an initial
state. This arises from computing requirement that registers
should be initialized to a known value before the start of
computation. Moreover, the initialization of qubits is used in
quantum error correction algorithms.

3. Long relevant decoherence times, much longer than the gate
operation time. It was shown that for fault-tolerant quantum
computation the magnitude of decoherence time scales should be
$10^4-10^5$ times the "clock time" of the quantum computer, that
is, the time for the execution of an individual quantum gate.

4. A 'universal' set of quantum gates. a) It should be possible to
perform precise unitary operations on the individual qubits. b)
Furthermore, the inter-qubit interaction should be controlled with
almost atomic precision.

5. A qubit-specific measurements capability. Any quantum computer
would deliver its output as results of measurements performed on
it.

In what follows we concentrate on nuclear spin based solid state
models \cite{PVK98,Kane98,Kane00,MPV01,MPG01,BMVW01,SSV01}. All
the existing models satisfy the criteria 1,2 and 4a. In some
models the criteria 3 is almost satisfied. At least, the ratio of
gate to decoherence time allows to create a few qubit quantum
computer. No existing model addresses the problem of controlling
the inter-qubit interaction with atomic precision. In this paper
we propose a new system in which criteria 1-4 are completely
fulfilled. Our investigation is mainly focused on the criterion
4b.

The proposed system consists of nuclear spins (qubits) embedded
into a zero nuclear spin mesoscopic ring or a finite length
quantum wire (Fig. 1a and Fig. 2a). The hyperfine interaction of
the electrons in the system with nuclear spins leads to an
effective indirect nuclear spin interaction. In what follows we
calculate the effective nuclear spin interaction energy. The
effective nuclear spin interaction energy can be chosen to have
the following form:

\begin{equation}
E=\left( I_{1,x}I_{2,x}+I_{1,y}I_{2,y}\right) A+I_{1,z}I_{2,z}B,
\label{inform}
\end{equation}
where $\overrightarrow{I_{i}}$ is magnetic moment of a nucleus,
$A$ and $B$ are functions of the system parameters as described
below. We obtain that the effective nuclear spins interaction
exhibits sharp maxima as function of the magnetic field and
nuclear spin positions which opens the way to manipulate qubits
with almost atomic precision. The selective nuclear spin
interaction can be obtained by changing external parameters of the
system.

The first calculation of the indirect, via electron spins, hyperfine
coupling between nuclear spins, was performed by Ruderman and Kittel \cite%
{Ruderman} for the case of 3D metal in the absence of magnetic field (see %
\cite{Slichter}). Influence of mesoscopic effects on the RKKY
interaction was studied by Spivak and Zyuzin \cite{Spivak}. They
pointed out that the only difference between pure and impure
metals is an additional random phase which depends on impurity
distribution. A mechanism of the indirect, via the exchange of
{\it \ virtual electron-hole} pairs (spin excitons), nuclear
spin interaction in the quantum Hall effect systems was suggested in \cite%
{Bychkov} and further elaborated in \cite{MPG01}. Quantum computation and
communication devices, based on this mechanism are proposed in \cite%
{PVK98,MPG01,SSV01}

\section*{Theoretical framework and results}

Let us consider a system consisting of electrons confined by a potential $%
V\left( {\bf r}\right)$ and interacting with two nuclear spins. We
assume that the nuclear spins are located far enough from each
other, so that the direct (dipole-dipole) nuclear spin interaction
is negligibly small as is in isotopically engineered Si/Ge
heterojunctions \cite{SSV01}. The contact hyperfine interaction
between electrons and nuclear spins leads to an indirect nuclear
spin interaction.

The wave function of the electron $\phi({\bf r})$ can be written
as a product of an envelope function $\Psi(\mathbf{r})$ by the
rapidly varying periodic function $u'_0(\mathbf{r})$ \cite{Paget}

\begin{equation}
\phi({\bf r})=\Psi({\bf r})u'_0({\bf r})=\Psi({\bf r})u_0({\bf r})
\left(V/\Omega \right)^{\frac{1}{2}} \label{Bloh}
\end{equation}
where $u_0(\mathbf{r})$ is the $\mathbf{k}=0$ Bloch state, $V$ is
the volume of the sample, and $\Omega$ is the volume of the unit
cell. The function $u'_0({\bf r})$ is conveniently normalized in
the cell volume $\Omega$, $\int \limits_{\Omega} \left| u'_0({\bf
r}) \right|^2d{\bf r}=1$. The localization of electrons is
described by the envelope part of the wave function $\Psi({\bf
r})$. The norm of the envelope function is $\int\limits_{V} \left|
\Psi({\bf r}) \right|^2d{\bf r}=\Omega$  \cite{Paget}.

In what follows we will consider the system in the envelope
function approximation, the rapidly varying function
$u'_0(\mathbf{r})$ appears only in the expression for hyperfine
interaction. The Hamiltonian of the system is given by $
H=H_{0}+H_{1}^{(1)}+H_{1}^{(2)}$ with
\begin{equation}
H_{0}=\frac{1}{2m^{*}}\left( {\bf p+}\frac{e}{c}{\bf A}\right)
^{2}+V\left( {\bf r}\right) -g\mu _{B}{\bf \sigma H} \label{Ho}
\end{equation}
and
\begin{equation}
H_{1}^{(i)}=\frac{8\pi }{3}\mu _{B}\hbar \gamma _{n} \left|
u'_0({\bf r}_i) \right|^2 {\bf \sigma I}_{i}\delta \left( {\bf
r}-{\bf r}_{i}\right) \label{H1}
\end{equation}
where $H_{0}$ is the Hamiltonian of the electron in the mesoscopic structure
in the magnetic field, $H_{1}^{(1)}+H_{1}^{(2)}$ is the perturbation due to
the contact hyperfine interaction, $m^{*}$ is the effective electron mass, $%
{\bf A}$ is the vector-potential, $g$ is the electron $g$-factor,
$\mu _{B}$ is the Bohr magneton, ${\bf H}$ is the magnetic field,
$\gamma _{n}$ is the nuclear gyromagnetic ratio, ${\bf I}_{i}$ and
${\bf \sigma }$ are nuclear and electron spins, ${\bf r}_{i}$ is
radius vector of $i$ nucleus, $i=1,2$. Because the electron wave
function is strongly peaked on the nuclei, the contact hyperfine
interaction energy greatly exceeds dipolar spin interactions.

The effective nuclear spin interaction energy calculated in a second-order
perturbation method is given by the expression \cite{Slichter}:

\begin{equation}
E=\sum_{E_{i},E_{f}}\frac{\left\langle \Psi _{i}\left| H_{1}^{(1)}\right|
\Psi _{f}\right\rangle \left\langle \Psi _{f}\left| H_{1}^{(2)}\right| \Psi
_{i}\right\rangle }{E_{f}-E_{i}}f_{i}\left( 1-f_{f}\right) +c.c.
\label{eff_en}
\end{equation}
Here $\Psi _{i}$ and $E_{i}$ are eigenfunctions and eigenvalues of
$H_{0}$ and $f_{i}$ is the electron distribution function. In this
Letter we restrict ourselves to the single electron approximation,
which proved to be
sufficient for clean low density quantum wires and rings, see \cite%
{HarrisonBk99} and references therein. Presence of impurities violates the
Kohn theorem \cite{Kohn61} and the electron interactions may play an
important role. In strongly correlated dense 1D electron systems the Fermi
liquid approach should be replaced by the Tomonaga-Luttinger theory, see %
\cite{AndoBk98} and references therein. The influence of the electron
correlations on the results obtained here will be the subject of a more
detailed publication.

\subsection*{Nuclear-spin interaction in mesoscopic rings}

Consider a torus-shaped quantum ring of inner radius $a$,
thickness $d\ll a$ and negligible height $h$ \cite{Alfons} in a
uniform parallel to the axis of the ring magnetic field ${\bf H}$
in $z$ direction with two nuclear spins located at ${\bf r}
(\rho,\varphi,z)= {\bf r}_1  \left( a+d/2, \varphi_1, h/2 \right)$
and ${\bf r}= {\bf r}_2 \left( a+d/2, \varphi_2, h/2 \right)$
(Fig.1a). In this subsection
we use the polar coordinates. The electron confining potential $%
V \left( \rho, z \right) $  is

\begin{equation}
V \left( \rho, z \right) = \cases{ 0 & if $a\leq \rho \leq a+d$
and $0\leq z \leq h$ ,
\\ \infty & otherwise.\\} \label{ro_potent}
\end{equation}

Firstly, let us find eigenfunctions and eigenvalues of the
Shr\"odinger equation with the Hamiltonian (\ref{Ho}). Due to the
axial symmetry of the ring, the wave function can be written as
follows

\begin{equation}
\Psi _{m,n,s=\pm }=\sqrt{ \frac{ \Omega }{2\pi}}\left( \left(
\begin{array}{c}
1 \\
0%
\end{array}
\right) ,\left(
\begin{array}{c}
0 \\
1%
\end{array}
\right) \right) e^{im\varphi }R_{m,n}\left( \rho \right)Z_1(z)
\label{Wavefunn}
\end{equation}
We assume that in $z$ direction the electron is always on the
ground level of the one-dimensional quantum well of thickness $h$,
thus $Z_1(z)=\sqrt{\frac{2}{h}}sin\left( \frac{\pi z}{h} \right)$.
Over the region $a\leq \rho \leq a+d$ the radial part of the wave
function $R_{m,n}\left( \rho \right)$ satisfies the equation:

\begin{equation}
-{\frac{\hbar ^{2}}{2m^{*}}\frac{1}{\rho }}{\frac{\partial }{\partial \rho }%
\rho }\frac{\partial }{\partial \rho }R_{m,n} +\frac{%
\hbar ^{2}}{2m_{e}\rho ^{2}}\left( m+\frac{\Phi (\rho )}{\Phi
_{0}}\right) ^{2}R_{m,n} =E_{m,n}R_{m,n} ,\label{radial_eq}
\end{equation}
where $\Phi _{0}$ is the magnetic flux quantum and the radial
number $n=1,2,3,...$ . The boundary conditions imposed by the
potential $V\left( \rho,z \right)$ (\ref{ro_potent}) are:
$R_{m,n}\left( a\right) =0$ and $R_{m,n}\left( a+d\right) =0$.

We will look for the solution of Eq.(\ref{radial_eq}) assuming that at $d\ll
a$ the vector-potential varies slow in the ring, so we can put $\Phi (\rho
)\simeq \Phi (a+d/2)$. In this case the solution of Eq.(\ref{radial_eq}) is
written in terms of Bessel functions:

\begin{equation}
R_{m,n}\left( \rho \right) =C_{1}J_{A}\left( \rho \alpha
_{m,n}\right) +C_{2}Y_{A}\left( \rho \alpha _{m,n}\right)
\label{rrr}
\end{equation}
where $A=\left| m+\frac{\Phi(a+d/2)}{\Phi_0} \right|$ and $\alpha
_{m,n}=\sqrt{\frac{2m_{e}E_{m,n}}{\hbar ^{2}}}$, the constants
$C_{1}$, $C_{2}$, energy levels $E_{m,n}$ are defined by boundary
conditions and the normalization condition
$\int\limits_{a}^{a+d}\rho R_{m,n}^{\ast }\left( \rho \right)
R_{m,n}\left( \rho \right) d\rho =1$. Unfortunately, the wave
function of the form (\ref{rrr}) allows to calculate the effective
nuclear spin interaction only numerically.

To obtain an analytical result, consider the effective interaction
energy constants $A$ and $B$ for the case of infinitely narrow
ring. To do this, let us set $\rho=a$ in (\ref{radial_eq}). It is
readily seen that in this case the radial part of wave function
decouples from the orbital part, $R_n( \rho)= \sqrt{ \frac{2} {d}}
sin \left( \frac{(r-a) \pi n}{d} \right)$ and

\begin{equation}
E_{n,m,s=\pm}=\frac{\hbar ^{2}}{2m^{*}} \left( \frac{\pi n }{d}
\right)^2 + \frac{\hbar ^{2}}{2m^{*}a^2} \left( m+\frac{\Phi
(a+d/2 )}{\Phi _{0}}\right) ^{2} \mp g\mu _{B}H/2.
\end{equation}

For the infinitely narrow ring it's reasonable to consider the
states only with $n=1$. With Eq.(\ref{eff_en}), the effective
nuclear spin interaction constants are:

\begin{equation}
A=K_r\sum_{m,n}\frac{\cos \left( \left(
n-m\right)\left( \varphi _{1}-\varphi _{2}\right) \right) }{E_{1,n,+}-E_{1,m,-}}f(E_{1,m,-})(1-f(E_{1,n,+}))%
\qquad ,  \label{EqA}
\end{equation}

\begin{equation}
B=K_r\sum_{m\neq n}\frac{\cos \left( \left(n-m\right) \left(
\varphi _{1}-\varphi _{2}\right) \right)
}{E_{1,n,+}-E_{1,m,+}}f(E_{1,m,+})(1-f(E_{1,n,+}))\qquad
\label{EqB}
\end{equation}
where $K_r=2\left( \frac{16}{3 a d h}\Omega \mu _{B} \hbar \gamma
_{n} \left| u'_0(0) \right|^2 \right) ^{2}$.

Using the wave function in the form Eq.(\ref{Wavefunn}), we
numerically calculate the magnetic field dependencies of the
nuclear spin interaction constants $A$ and $B$ using the following
set of parameters: $\varphi _{1}=0$, $\varphi _{2}=\pi$,
$a=100nm$, $h=1nm$, $d=1nm$ and $0.5nm$, one electron in the ring,
$T=0$ . The results of our calculations are shown on Fig.1b. The
magnetic field dependence of the effective nuclear spin
interaction constants are defined by the energy level statistics.
The nuclear spin interaction constant $A$ describes nuclear spin
flip-flop processes which are performed through flips of electron
spin, and, therefore, the main contribution to this process is due
to the energy levels with the same orbital quantum numbers and
different spin directions, what gives $1/H$ dependence of the
interaction amplitude.  The nuclear spin coupling constant $B$ is
connected with electron transitions between energy levels with
different orbital quantum numbers, but the same electron spin
direction. Their periodicity with the magnetic field results in
the periodicity of the constant $B$. It is seen that at the values
of the magnetic field, when there are two ground states, function
$B$ has discontinuities. In this case the perturbation theory is
not applicable. The obtained result is in a good agreement with
the result obtained for the case of infinitely narrow ring
potential Eqs.(\ref{EqA},\ref{EqB}).

\subsection*{Nuclear-spin interaction in mesoscopic wires}

The next system under consideration is a finite length quantum
wire of the length $l$ in $x$-direction, of the thickness $d$ and
of the height $h$ with
two nuclear spins located at $%
{\bf r}(x,y,z)={\bf r}_{1}\left( l_{1},d/2,h/2\right) $ and $ {\bf
r}={\bf r}_{2}\left( l_{2},d/2,h/2\right) $ in a magnetic field
${\bf H}$ in $z$ direction (Fig. 2a). We suppose that the
transversal sizes of quantum wire are much smaller than the length
of the quantum wire and the cyclotron orbit of electron. We
consider a model, when the confining potential is

\begin{equation}
V(x) = \cases{ 0 & if  $0\leq x\leq l$, $0\leq y\leq d$, $0\leq
z\leq h$\\ \infty & otherwise.\\} \label{pottt}
\end{equation}
The eigenfunctions and eigenvalues of the Hamiltonian (\ref{Ho})
with potenital (\ref{pottt}) are

\begin{equation}
\Psi _{n,m,k,\pm }=\sqrt{\frac{8}{ldh}}\left( \left(
\begin{array}{c}
1 \\
0%
\end{array}
\right) ,\left(
\begin{array}{c}
0 \\
1%
\end{array}
\right) \right) \sin \left( \frac{n\pi x}{l}\right)  \sin \left(
\frac{m \pi y}{d}\right)  \sin \left( \frac{k \pi z}{h}\right),
\label{wavef1}
\end{equation}

\begin{equation}
E_{n,m,k,\pm }=\frac{\hbar ^{2}\pi
^{2}}{2m^{*}l^{2}}n^{2}+\frac{\hbar ^{2}\pi
^{2}}{2m^{*}d^{2}}m^{2}+\frac{\hbar ^{2}\pi
^{2}}{2m^{*}h^{2}}k^{2}\mp g\mu _{B}H/2, \label{enlevel}
\end{equation}
As in the case of the nuclear spin interaction in mesoscopic
wires, we assume that $d,h\ll l$, so we consider only the
electrons on the ground levels of potential wells in $y$ and $z$
directions, i.e. $m=k=1$. Substitution of Eqs. (\ref{wavef1}) and
(\ref{enlevel}) into (\ref{eff_en}) gives us the effective nuclear
spin interaction constants.

We have calculated analytically and numerically the effective
interaction constants $A$ and $B$ as a functions of nuclear spin
positions for odd and even number of electrons $N$ in the wire.
The results of calculation for the odd number of electrons ($N=9$)
at $T=0$ are presented at Figures 2b-2d. In low magnetic field
region, when Zeeman splitting energy is much less than the energy
gap between levels with different\ $n$ (Eq.(\ref{enlevel})) and at
$T=0$ limit the expression for $A$ takes a simple fortm:

\begin{equation}
A=K \frac{\left( \sin \left(
\frac{(N+1)\pi l_{1}}{2l}\right) \sin \left( \frac{(N+1)\pi l_{2}}{2l}%
\right) \right) ^{2}}{g\mu _{B}H}. \label{coeA}
\end{equation}
where $K_w=2\left( \frac{64 \pi \Omega \mu _{B}\hbar \gamma _{n}
\left|u'_0(0) \right|^2 }{3ldh}\right) ^{2}$. In this limit the interaction constant $%
A $ has a set of $\frac{(N+1)^2}{4}$ maximums (Fig. 2b) and $B\ll
A$. Fig. 2c shows an increasing of interaction constant $A$ if
nuclear spins are located not far from each other (in the vicinity
of the line $l_{1}=l_{2}$) and a decreasing of interaction
constant $A$ for the other nuclear spin positions with increasing
of the magnetic field. The interaction constant $B$ has a
non-trivial dependence on the nuclear spin positions (Fig. 2d).

\section*{Conclusions and discussion}

To conclude, we proposed a new possible implementation of a basic
unit for quantum computer based on the nuclear-spin qubits
embedded into the zero nuclear spin mesoscopic ring or finite
length quantum wire. Particular emphasis has been placed on the
investigation of nuclear spin interaction via the electrons
confined in such systems. It was found that the indirect
nuclear-spin qubit interaction is very sensitive to the system
parameters: nuclear spin location, number of electrons, magnetic
field and geometry of the system. It dependence on the system
parameters is completely different from indirect nuclear spin
interaction in 2D and 3D metals. Preliminary finite temperature
calculation indicates that the values of the effective nuclear
spin interaction constants are decreasing with temperature, but
the main features of the obtained results remain qualitatively
unchanged.

Now let us consider how our model of quantum computer is matched
by the set of DiVincenzo criteria listed in the Introduction.

1. It's well known that nuclear spins are appropriate candidates
to be qubits \cite{Kane98,Privman02}. Nuclei with spin $1/2$ are
two-level systems with well defined states $|0>$ and $|1>$.
Needless to say that such a system is scalable and, basically,
there is no principal limitation on the reasonable number of
nuclear-spin qubits which can be integrated into a quantum
circuit. The ring architecture of the quantum computer is of
considerable promise. Let us imagine a mesoscopic ring with
nuclear-spin qubits located in the immediate vicinity of it. A
local change of the electrostatic potential near a qubit by a gate
electrode changes the value of the envelope wave function on the
qubit and, correspondingly, allows to switch on the interaction
between any two qubits, whereas almost at all quantum computer
proposal only adjacent qubits can directly interact. Nuclear-spin
qubit interaction through electrons confined on a sphere opens a
further way to improve the quantum computer architecture.

2. We propose to initialize the nuclear-spin qubits using
spin-polarized electrons. There is a general agreement that it is
the only feasible method since the nuclear spins are highly
isolated from the environment. Possibility of nuclear spin
polarization by spin-polarized transport was demonstrated almost
ten years ago \cite{Wald,Kronm99}. Possible methods to introduce
nonequilibrium polarizations of the electrons include injections
of spins from ferromagnetic contacts \cite{Prinz}, optical pumping
\cite{BTPW94,Barrett99b} and spin refrigeration (see \cite{Kane00}
and references therein).

3. The decoherence time of the nuclear spins in mesoscopic systems
is expected to be long enough to perform the quantum computation,
since the discrete electron spectrum in mesoscopic systems imposes
restriction on the flip-flop processes and the nuclear spin
relaxation time at law temperatures is expected to have an
activation behavior \cite{VM88}. Up to the present, a calculation
of nuclear spin decoherence time in mesoscopic structures is not
made, however as a rough estimation we can take the decoherence
time of the nuclear-spin qubit embedded into 2DEG, $T_2=10$ sec
\cite{Privman02}. The characteristic time of qubit interaction is
$T_{int}\sim h/A$, where for the nuclear-spin qubits in finite
length quantum wire $A$ is given by Eq. (\ref{coeA}). Using
available experimental data for $GaAs$, $\left| u'_0(0)
\right|^2_{^{75}As}=9.8\cdot 10^{25}cm^{-3}$ \cite{Paget} and
following set of parameters: $a=200nm$, $d=h=5nm$, $H=0.01T$ we
obtain $T_{int}=2 \cdot 10^{-5}sec$ and $T_2/T_{int}=5 \cdot
10^{5}$. Our estimate indicates that the present system is
suitable for fault-tolerant quantum computation.

4. a) The one-qubit operations using the NMR could be similar to
the existing experimental suggestions \cite{PVK98,MPG01,SSV01}. b)
Interaction between any two qubits, which is necessary for
two-qubit operations, is performed by the confined electrons. By
varying the external parameters (magnetic field, number of
electrons, gate potentials) we can control with almost atomic
precision the nuclear spin interaction strength by creating maxima
of the amplitude of electron wave function on some qubits and zero
on the other. As an example, consider how it is possible to
control nuclear-spin qubit interaction in the finite length wire
by changing the number of electrons.  Let us place the qubits at
some predefined positions with coordinates
$x_i=\frac{\alpha_i}{\beta_i}$, where $\beta_i$ are different
prime numbers and $\alpha_i$ are integers ($\alpha_i<\beta_i$).
The orbital part of the last electron level wave function (from
Eq.(\ref{wavef1})) for

\begin{equation}
N_{jk}=2\frac{\prod \limits _{i=1}^M \beta_i}{\beta_j \beta_k}-1,
\end{equation}
electrons in the system is

\begin{equation}
\Psi _{n,m,k,\pm }=\sqrt{\frac{8}{ldh}} \sin \left(\frac{ \pi
\prod \limits _{i=1}^M \beta_i}{l \beta_j \beta_k} x \right)  \sin
\left( \frac{m \pi y}{d}\right)  \sin \left( \frac{k \pi
z}{h}\right). \label{wavef2}
\end{equation}
It is readily seen that $\Psi _{n,m,k,\pm } (x_i)=0$ for $i\neq
j,k$. This means, that all qubits except $j$ and $k$ are at the
nodes of the wave function and only $j$ and $k$ qubits interact.
Fig. 3 illustrates this idea. It follows that the accuracy of
nuclear spin positioning should be few atomic units.

5. In mesoscopic systems the single nuclear-spin measurement is
still an open problem. One of the possibilities is to use
Hyperfine Aharonov-Bohm effect (HABE) \cite{VRWZ98}, following
from the coupling of the nuclear spin polarization to the phase of
the conduction electron wave function. This was outlined in
\cite{PVK98} and will be considered in details elsewhere. Another
possibility is to use spin-dependent magneto-transport tunneling
through a system, which energy levels are spin-split by an
external magnetic field \cite{Fedya}. The tunneling current
develops a distinctive peak at the frequency of Zeeman splitting,
which can be sufficiently narrow to measure a state of the nuclear
spin. Moreover, it's possible to use spin to charge conversion as
it was discussed in \cite{Kane00}.

\section*{Acknowledgments}

We gratefully acknowledge helpful discussions with Yu. A. Bychkov,
A.Dyugaev, T. Maniv, D. Mozyrsky, V. Privman, S. Safarov and I.
Shlimak. This research was supported in part by the 5th European
Program, IST 2000 29686 NSP-SI, by the Israel Science Foundation,
by the National Science Foundation, grants DMR-0121146 and
ECS-0102500, and by the National Security Agency and Advanced
Research and Development Activity under Army Research Office
contracts DAAD-19-99-1-0342 and DAAD 19-02-1-0035.

\newpage

\begin{figure}
\centerline{ \psfig{figure=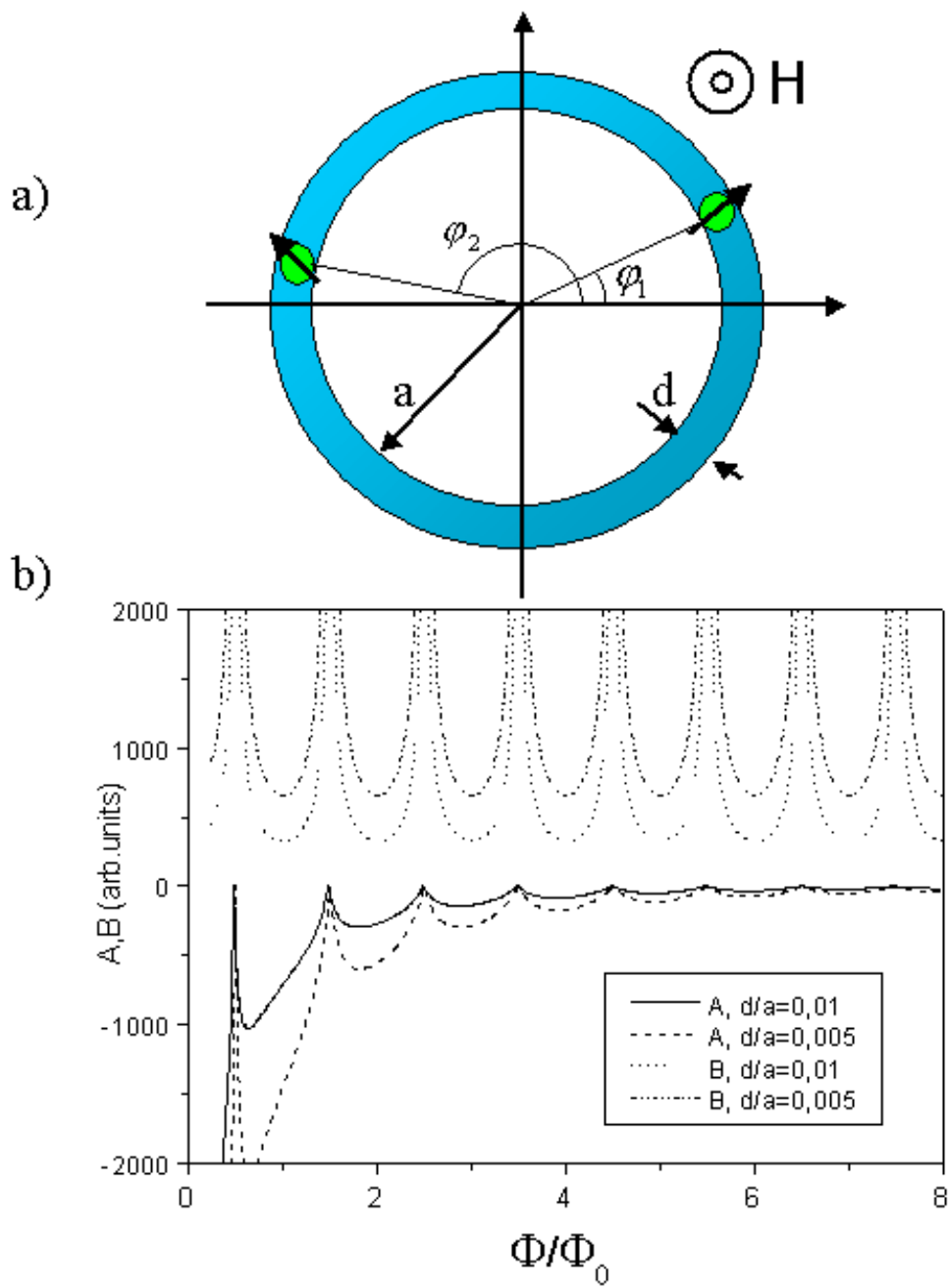,width=15cm}}
 \caption{ a) Two
nuclear spins embedded into a torus shaped quantum ring; b)
dependence of nuclear spin coupling constants $A$ and $B$ on the
magnetic field, $E_r=\frac{\hbar^2}{2m^*a^2}$.}
\end{figure}

\begin{figure}
\centerline{ \psfig{figure=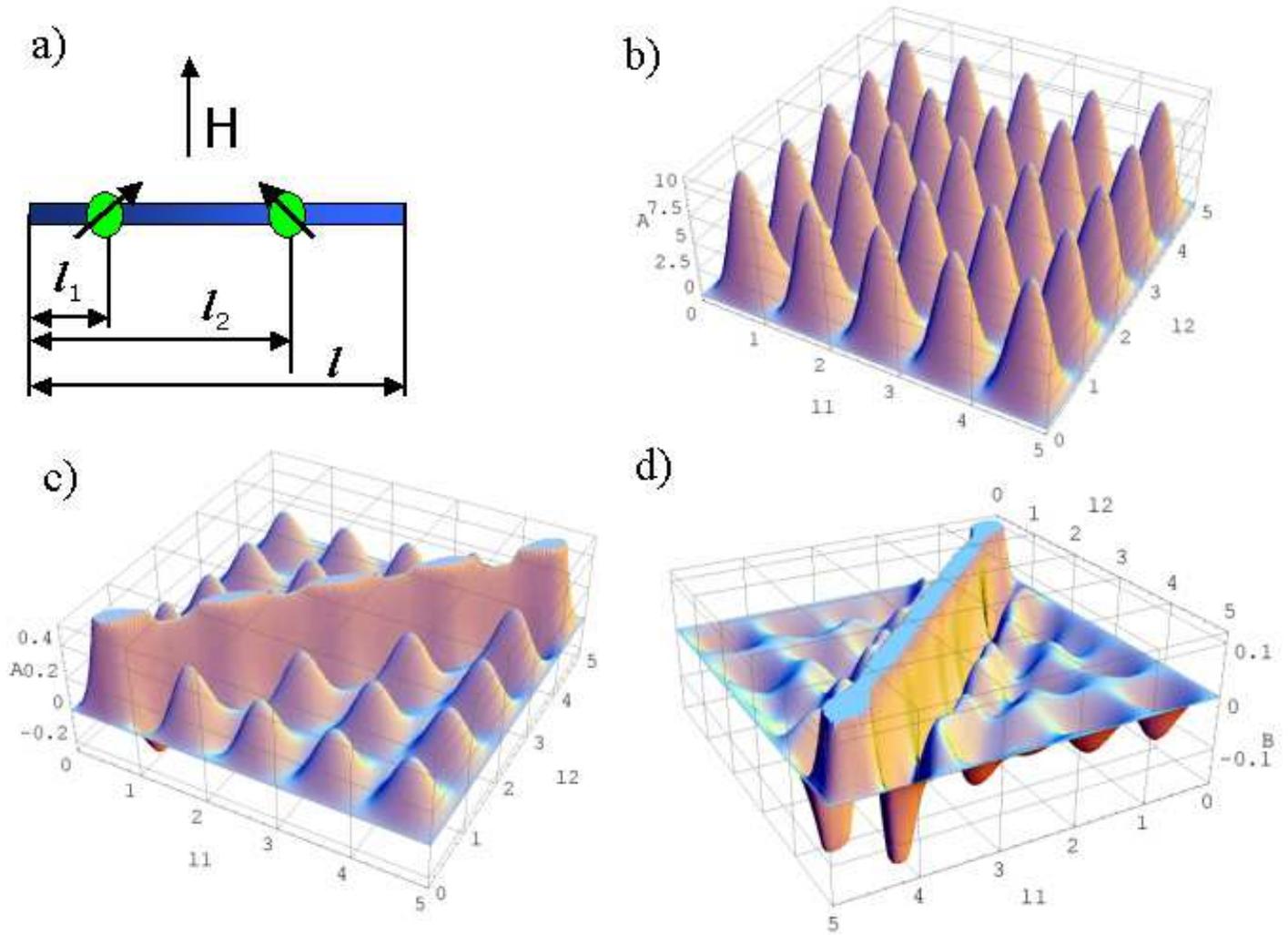,height=15cm,angle=270}}
 \caption{a) Two nuclear spins embedded into a finite length quantum wire in a
magnetic field; b) nuclear spin coupling constant $A$ in low
magnetic field region ($\frac{4g\mu _{B}m_{e}a^{2}H}{\hbar ^{2}\pi
^{2}}=0.1$); c) nuclear spin coupling constant $A$ in high
magnetic field region ($\frac{4g\mu _{B}m_{e}a^{2}H}{\hbar ^{2}\pi
^{2}}=3$) and d) nuclear spin coupling constant $B$ on the nuclear
spin positions. Odd number of electrons, $T=0$, the nuclear spin
coupling constants are given in units of $K_w/E_w$, where
$E_w=\frac{\hbar^2 \pi^2}{2m^*l^2a  }$ .}
\end{figure}

\begin{figure}
\centerline{ \psfig{figure=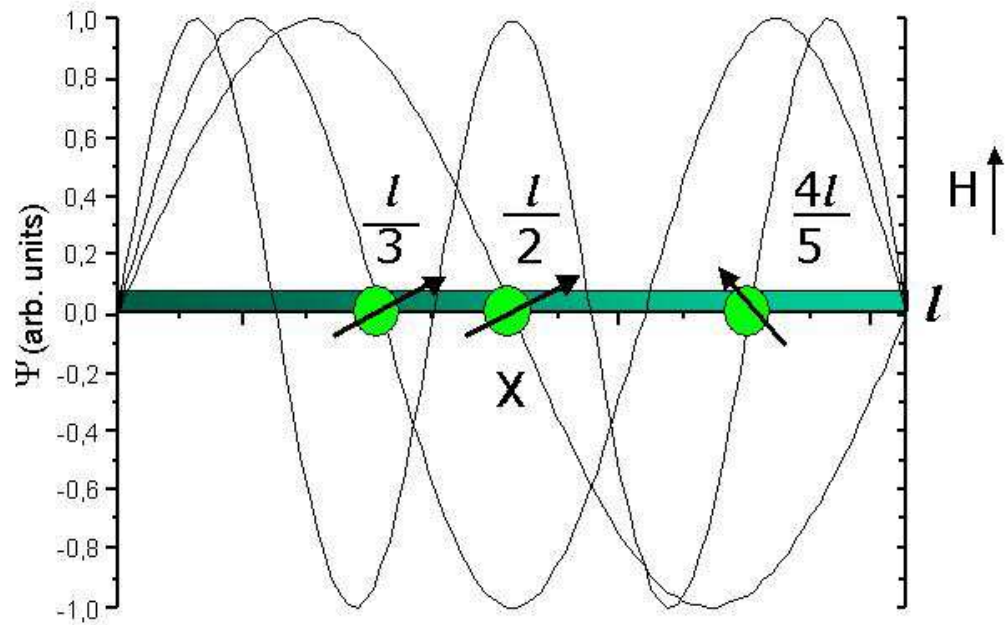,height=12cm,angle=270}}
 \caption{Qubit
arrangement and the last occupied electron level wave function for
$N=3,5,9$ electrons in the wire. The qubits are located in the
nodes of the different wave functions.}
\end{figure}

\end{document}